\pdfoutput=1
\documentclass[12pt,a4paper]{article}
\usepackage{xcolor}
\usepackage{enumerate}
\usepackage{framed}
\usepackage{mdframed}
\usepackage{amsmath}
\usepackage{amsfonts}
\usepackage{mathrsfs}
\usepackage{amssymb}
\usepackage{graphicx, rotating}
\usepackage{epsfig}
\usepackage{latexsym}
\usepackage{graphicx}
\usepackage{color}
\usepackage{amsmath,bm,amssymb}
\usepackage{cite}

\usepackage{slashed}
\usepackage{hyperref}
\usepackage{epstopdf}
\usepackage[font=footnotesize,labelfont=]{caption}
\AppendGraphicsExtensions{.tif}
\hypersetup{colorlinks=true, citecolor=bluscuro, linkcolor=black, urlcolor=bluscuro}
\definecolor{rossos}{cmyk}{0,1,1,0.55}
\definecolor{bluscuro}{rgb}{0.15, 0.2, .85}
\definecolor{bluchiaro}{cmyk}{1,.3,0.,0.1}

\def\0{\vec{0}}

\def\t{\eta}

\def\del{\partial}

\def\d{{\rm d}}
\def\vx{{\vec{x}}}

\def\vk{{\vec{k}}}

\def\t{\tau}
\def\s{\sigma}
\def\a{\alpha}
\def\Bl{\Big{(}}
\def\Br{\Big{)}}
\def\bl{\big{(}}
\def\br{\big{)}}

\def\beq{\begin{equation}}
\def\eeq{\end{equation}}

\begin{document}
\def\thefootnote{\fnsymbol{footnote}}

\begin{center}
\Large{\textbf{On  the Inflationary Perturbations of \\  Massive Higher-Spin Fields   }} \\[0.5cm]
\end{center}
\vspace{0.5cm}

\begin{center}

\large{Alex Kehagias$^{\rm a,b}$  and  Antonio Riotto$^{\rm c}$}
\\[0.5cm]

\small{
\textit{$^{\rm a}$Physics Division, National Technical University of Athens, 15780 Zografou Campus, Athens, Greece}}

\vspace{.2cm}

\small{
\textit{$^{\rm b}$Theoretical Physics Department, CERN, CH-1211 Geneva 23, Switzerland}}

\vspace{.2cm}

\small{
\textit{$^{\rm c}$Department of Theoretical Physics and Center for Astroparticle Physics (CAP) \\
24 quai E. Ansermet, CH-1211 Geneva 4, Switzerland}}

\vspace{.2cm}

\end{center}

\vspace{.7cm}

\hrule \vspace{0.3cm}
\noindent \small{\textbf{Abstract}\\ 
Cosmological  perturbations of  massive higher-spin fields  are generated during inflation, but they decay on scales larger than the Hubble radius as a consequence of the Higuchi bound. By introducing  suitable couplings to  the inflaton field, we show that   one can obtain statistical correlators
of  massive higher-spin fields which remain constant or decay very slowly outside the Hubble radius. This opens up the possibility of new observational 
signatures from inflation.}

\vspace{0.3cm}
\noindent
\hrule
\def\thefootnote{\arabic{footnote}}
\setcounter{footnote}{0}

%
%
%
%


 \def\vx{\vec{ x}} 
\def\vk{\vec{k}}
\def\vy{\vec{y}}

\numberwithin{equation}{section}

\def\la{~\mbox{\raisebox{-.6ex}{$\stackrel{<}{\sim}$}}~}
\def\ga{~\mbox{\raisebox{-.6ex}{$\stackrel{>}{\sim}$}}~}
\def\bq{\begin{quote}}
\def\eq{\end{quote}}
\def\PL{{ \it Phys. Lett.} }
\def\PRL{{\it Phys. Rev. Lett.} }
\def\NP{{\it Nucl. Phys.} }
\def\PR{{\it Phys. Rev.} }
\def\MPL{{\it Mod. Phys. Lett.} }
\def\IJMP{{\it Int. J. Mod .Phys.} }
\font\tinynk=cmr6 at 10truept
\newcommand{\be}{\begin{eqnarray}}
\newcommand{\ee}{\end{eqnarray}}
\newcommand{\n}{{\bf n}}
\newcommand{\arXiv}[2]{\href{http://arxiv.org/pdf/#1}{{\tt [#2/#1]}}}
\newcommand{\arXivold}[1]{\href{http://arxiv.org/pdf/#1}{{\tt [#1]}}}

\section{Introduction \label{sec:intro}} 
The problem of writing down consistent    equations of motion and Lagrangians  for higher-spin (HS)  fields goes back to the beginning of quantum field theory (for reviews, see Refs. \cite{taronna,giombi}) and  is particularly difficult for massless fields. Massless degrees of freedom with spin $s\geq 1$ are gauge fields and  they  come with the  corresponding  gauge invariance
 needed to decouple unphysical polarizations. The problem of writing consistent self-interactions become therefore  highly constrained and complicated.
 
  In flat space one can write down consistent gauge-invariant equations of motion for the free fields, but it seems impossible to have non trivial $S$-matrices for spins $s>2$ since the gauge invariances are accompanied with conserved charges and the conservation laws are too strong to allow non-trivial
$S$-matrices. This is consistent with Coleman-Mandula theorem stating that the $S$-matrix in flat spacetime cannot have extra symmetries beyond the (super-)Poincar\' e symmetry. On the other hand, there are some explicit constructions of self-interacting massless HS  theories away from flat spacetime when a non-vanishing cosmological constant is allowed and no $S$-matrix exists \cite{vas}. This is particularly interesting when thinking of the possible role of HS fields during inflation. These theories   always contain the massless spin-2 graviton and are therefore theories of gravity.  An important  feature of these HS theories  is that their mathematical consistency implies that they involve an infinite tower of fields of all spins.

 Even though the existence of such theories
may look surprising given the large number of constraints, they look natural from the AdS/CFT \cite{ads}, or dS/CFT \cite{strom},  point of view  and the subject of an intense research activity, see Refs. \cite{sthesis,s1} and references therein. For instance, one can compute  the cubic couplings of  the minimal bosonic HS theory  in AdS$_{4}$
starting from the holographic dual theory \cite{s2}. Of course, to assess  the  importance of massless HS fields during inflation one has to
deal not only with the infinite tower of degrees of freedom (which might turn out to be a bonus from the  observational point of view), but also to compute 
the couplings of the massless HS fields to the matter (inflaton) sector. This calculation, better performed on the CFT side, will allow a reliable computation of the statistical inflationary correlators and will be presented elsewhere \cite{inprep}.

In this paper we take a more modest approach and deal with massive HS spins during inflation. Their signatures on the non-gaussian cosmological correlators of the comoving curvature perturbation 
have been recently studied in Ref. \cite{am} (see also Ref. \cite{baumann}). They arise in the squeezed limit of the correlation functions when intermediate  HS fields are exchanged carrying  informations about their masses and    spins.
If measured, these imprints will  provide  an exciting information about the particle spectrum during  the inflationary universe \cite{chen}.  However, despite the fact that gauge invariance
does not constrain the system so tightly, the de Sitter isometries impose the so-called  Higuchi bound \cite{hig} on the masses of the HS states, 

\beq
m^2>s(s-1)H^2,
\eeq
 where $H$ is the Hubble rate during inflation. This bound, which has a neat interpretation if derived from the 
 CFT$_3$ side of the dS/CFT correspondence \cite{am,cr,us},  implies the  absence of curly hair in de Sitter \cite{cr}. On wavelengths larger than the Hubble radius the perturbations of the fields with spin  $s$ are scaling as a function of the conformal time  $\tau$ as  $(-\tau)^\Delta$, where 
 
 \beq
 \frac{m^2}{H^2}=-\Delta(\Delta-3)+(s-2)(s+1).
 \eeq
The Higuchi bound imposes $\Delta>1$ and  HS fluctuations are doomed to promptly decay as soon as they leave the Hubble radius.  As such, the HS fields are short-lived  mediators and the corresponding signatures in the four-point correlator of the curvature perturbation  are  suppressed by powers of the exchanged momentum
in the squeezed configuration.

On the other hand, it is well-known that one can obtain vector spin-1 perturbations which remain constant of super-Hubble scales  by modifying the
kinetic term to $I(\phi)F_{\mu\nu}^2$, that is by introducing  
 an appropriate function of time (or equivalently inflaton field $\phi$) \cite{v1,v2}. In such a case, one can  characterize the correlators involving the inflaton and the vector fields by exploiting the fact that 
the de Sitter isometry group acts as conformal group on the three-dimensional Euclidean space \cite{vector}.

The goal of this paper is to extend to generic HS fields what is known for vectors and to investigate what kind of time-dependent  functions one needs to couple the HS fields to in order to generate  correlation functions which can decay slower than what dictated by the Higuchi bound outside the Hubble radius. 

We will follow a bottom-up approach and 
start from the equation of motion of the HS fields.  We will see that the requirement of having the correct number of propagating degrees of freedom drastically reduces the   possible choices of the functions as well as the way they couple  to the HS fields. 
For some choice within the allowed set of   functions
the HS perturbations remain  constant on scales larger than the Hubble radius   and an  enhanced symmetry shows up. For some cases, we will also  be able to derive the corresponding actions. Long-lived HS fluctuations may not only  leave a seizable imprint on the statistical correlators of the scalar perturbations as intermediate states, but also alter the dynamics of scalar and  tensor  perturbations and possibly give rise to detectable observables with HS fields on the external legs. 

The paper is organized as follows. In section 2 we analyze the case of the spin-1, which is the most known in the literature. In particular we show that one can recover the known result by simply starting at the level of the most general equation of motion, instead from the action. Section 3 is devoted to the study of the
spin-2 fields. We will write the most generic equation and constraints, derive the orthonormality condition, the corresponding  
Higuchi bound and discuss the cases in which extra gauge symmetries appear in the system. We will also identify for which suitable coupling to the inflaton there exist perturbations of the helicities $\pm 2$ which remain constant on super-Hubble scales. 
The case of the generic spin-$s$ is discussed in Section 4. Section 5 is devoted to a short descriptions of possible
observational consequences. Finally, we briefly conclude in Section 6.

\section{The spin-1 case}
We start our analysis with the simplest case of the vector field dynamics during a de Sitter phase with spacetime metric 

\begin{eqnarray}
\d s^2
=\frac{1}{H^2 \t^2}\left(-\d \t^2+\d \vec{x}^2\right).
\label{metric}
\end{eqnarray}
Here $H$  is the Hubble rate and we imagine that the inflationary phase is driven by a scalar inflaton field 
whose vacuum expectation value $\phi_0(\t)$ is slowly varying with time in such a way that the background metric can be approximated by the expression (\ref{metric}).  We wish to understand if  it possible to couple  a spin-1 field suitably to a function of the inflaton in such a way that its helicities $\pm 1$ of the canonically normalized super-Hubble perturbations can stay constant in time\footnote{From the helicity equations
written later on, it is easy to show that, given a spin-$s$ state, if the helicities $\pm s$ are constant on super-Hubble scales, all the other helicities  decay on large-scales with increasing powers of the conformal time.}.

A   spin-1 field $\sigma^\rho$ on the de Sitter background with mass 
$m$ satisfies the following equation 
\begin{eqnarray}
\left(\square -m_1^2\right)\sigma^\rho=0, 
\end{eqnarray}
where 
\begin{eqnarray}
m_1^2=m^2+3 H^2.
\end{eqnarray}
We now couple the  spin-1 field $\sigma^\rho$ to functions of time, which we might think  of as functions of $\phi_0$. 
The most general coupling up to two derivatives 
 is of the form

\begin{eqnarray}
\square\sigma^\rho+\left(\nabla^\mu I\right) \nabla_\mu \sigma^\rho+
\alpha \nabla_\mu I \nabla^\rho \sigma^\mu+\beta 
I^{\rho}_\mu\sigma^\mu-M_1^2(\phi)\sigma^\rho=0.
  \label{gen1}
\end{eqnarray}
The constraint 
\begin{eqnarray}
\nabla_\rho\sigma^\rho=0,  \label{co1}
\end{eqnarray}
 ensures that Eq. (\ref{gen1}) for $\sigma^\rho$ propagates three-degrees of freedom, that is the degrees of freedom expected for a massive spin-1 field.  
The parameter $\beta$, the form factors $I(\phi)$, $M_1^2(\phi)$ and ${I^\rho}_\mu(\phi)$ are  restricted  by the consistency of the equation with the constraint. Taking the divergence of  equation of motion and using Eq. (\ref{com}),
 we find 
\begin{eqnarray}
&&\left(\nabla_\rho\nabla^\mu I\right)\nabla_\mu \sigma^\rho+3H^2 \nabla_\mu \sigma^\mu+\alpha (\nabla_\rho\nabla_\mu I)\nabla^\rho \sigma^\mu-\alpha
\nabla_\mu I\nabla^\nu I \nabla_\nu \sigma^\mu\nonumber \\
&&-\alpha^2\nabla_\mu I \nabla_\nu I \nabla^\mu \sigma^\nu -
\alpha \beta \nabla_\mu {I^\mu}_\nu \sigma^\nu +\alpha M_1^2 \nabla_\mu I\,  \sigma^\mu 
\nonumber \\
&&+
\beta\left(\nabla_\rho{I^\rho}_\mu\sigma^\mu\right)
+\beta{I^\mu}_\rho \nabla_\mu \sigma^\rho
-\left(\nabla_\rho M_1^2\right)\sigma^\rho=0, 
\end{eqnarray}
which can be written as 

\begin{eqnarray}
0&=&\nabla_\mu \sigma^\rho\Big{\{}\beta {I^\mu}_\rho+(1+\a)(\nabla_\rho\nabla^\mu I-\a \nabla^\mu I \nabla_\rho I)\Big{\}}\nonumber \\
&&+\sigma^\rho\Big{\{}
(3H^2+\alpha M_1^2)\nabla_\rho I+\beta \left(\nabla_\mu {I^\mu}_\rho-\alpha {I^\mu}_\rho\nabla_\mu I \right)-\nabla_\rho M_1^2\Big{\}}.
\end{eqnarray}
This equation is satisfied without imposing any further constraint on $\sigma^\mu$ for 
\begin{eqnarray}
\beta I_{\mu\nu}=-(\alpha+1)\left(\nabla_\mu\nabla_\nu I-\alpha 
\nabla_\mu I \nabla _\nu I\right)  \label{i10}
\end{eqnarray}
and 
\begin{eqnarray}
(3H^2+\alpha M_1^2)\nabla_\mu I+\beta \left(\nabla_\rho {I^\rho}_\mu-\alpha {I^\rho}_\mu\nabla_\rho I \right)-\nabla_\mu M_1^2=0. \label{m0m}
\end{eqnarray}
Therefore, the functions $M_1^2$ and  ${I^\mu}_\nu$ are determined by Eqs. (\ref{i10}) and (\ref{m0m}), once the function $I(\phi)$ is  specified. 
The equation obeyed by $\sigma_\mu$ is explicitly written as 

\begin{eqnarray}
\sigma_i''-\nabla^2\sigma_i-\frac{2}{\t} \partial_i\sigma_\t+I'(\sigma_i'+\a\partial_i \sigma_\t)+\frac{M_1^2/H^2-3}{\t^2}\sigma_i+2(1+\alpha)\frac{I'}{\t}\sigma_i&=&0\, \\
\sigma_\t''-\nabla^2\sigma_\t-\frac{2}{\t}\partial_i\sigma_i+\frac{M_1^2/H^2-1}{\t^2}\sigma_\t-(1+\alpha)\left(I''-a{I'}^2\right)\sigma_\t+(1+\a)I'\s_\t'&=&0, \\
\sigma_\t'-\frac{2}{\t}\sigma_\t&=&\partial_i\sigma_i.  \label{gc1}
\end{eqnarray}
\subsection{Long-lived spin-1 perturbations}
In order to obtain scaling solutions, we may choose $I$ to be of the form 

\begin{eqnarray}
I=n\ln(-H\tau)=n\int^\phi{\rm d}\phi'\frac{V(\phi')}{V'(\phi')}. \label{I}
\end{eqnarray}
  Then the solution of Eq. (\ref{m0m}) with a constant mass turns out to be
\begin{eqnarray}
M_1^2=\Big{[}3-n(1+\a)(3+\a\, n)\Big{]}H^2. \label{M12}
\end{eqnarray}
We may now expand $\sigma_\mu$ in helicity modes as 
\begin{eqnarray}
\sigma_\mu=\sum_{\lambda=-1}^1\sigma^{(\lambda)}_{\mu},
\end{eqnarray}
where 
\begin{eqnarray}
\sigma^{(0)}_\t=\sigma^0_{0,1},&&\sigma^{(\pm1)}_\t=0,\nonumber \\
\sigma^{(0)}_i=\sigma^0_{1,1}\varepsilon^0_i,&&\sigma^{(\pm1)}_i
=\sigma^{(\pm 1)}_{1,1}\varepsilon^{\pm1}_i,
\end{eqnarray}
and  the polarization vectors $\varepsilon^\lambda_i$ are normalized as 
\begin{eqnarray}
\hat{k}_i \varepsilon^0_i=1, ~~~\hat{k}_i\varepsilon^{\pm 1}_i=0, ~~~\varepsilon^{\pm 1}_i\varepsilon^{\mp1}_i=2,~~~ \varepsilon^{\mp 1}_i=\varepsilon^{\pm 1 *}_i. 
\end{eqnarray}
These conditions are solved for $\varepsilon^0_i=\hat{k}_i$ and  for momentum along the $z$-direction, we may choose $\varepsilon^{\pm 1}_i=(1,\pm i,0)$. 
The equations for the  helicity modes become
\begin{eqnarray}
{\sigma^{\pm 1}_{1,1}}''+\frac{n}{\t}{\sigma^{\pm 1}_{1,1}}'+\left(k^2-\frac{n(1+\a)(1+\a\, n)}{\t^2}\right)\sigma^{\pm 1}_{1,1}&=&0, \label{pp1}\\
{\sigma^{0}_{0,1}}''-\frac{2-n(1+\a)}{\t}{\sigma^{0}_{0,1}}'+\left(k^2+\frac{2\left[1-n(1+\a)\right]}{\t^2}\right)\sigma^{0}_{0,1}&=&0. \label{pp2}
\end{eqnarray}
The longitudinal mode $ \sigma^0_{1,1}$ is specified by Eq.(\ref{gc1}) to be
\begin{eqnarray}
\sigma^0_{1,1}=-\frac{1}{k}\left({\sigma^0_{0,1}}'-\frac{2}{\t}\sigma^0_{0,1}\right). \label{s001}
\end{eqnarray}
The equation for obeyed by the longitudinal mode $ \sigma^0_{1,1}$ 
can easily be found by appropriate differentiation of the (\ref{s001}). 
For a field scaling as 
\begin{eqnarray}
\sigma_\mu(\tau,\vec{x})=\tau^{\Delta-1} \widetilde\sigma_\mu(\vec{x}), \label{ss}
\end{eqnarray}
we find  

\begin{eqnarray}
\Delta_-= 2+n \, \a ~~~{\rm and}~~~ \Delta_+=1-(1+\a)n. \label{cd}
\end{eqnarray}
Let us  now consider  the canonically normalized field (from Eq. (\ref{gen1}) one can see that  this choice combines the first two terms to give only $\square \bar{\sigma}_{i}$ plus other interaction pieces)

\be
\bar{\sigma}_i=\Big<{\rm exp}(I(\phi)/2)\Big>\sigma_i={\rm exp}(I(\phi_0)/2)\sigma_i. \label{nor}
\ee
This field has   scaling behaviour $\bar\Delta-1$ and  from  Eq. (\ref{cd}) we infer

\be
\bar\Delta-1=\Delta-1+\frac{n}{2},
\ee
so that 

\be
\bar\Delta_-=2+\frac{n}{2}(1+2\a), ~~~\bar\Delta_+=1-\frac{n}{2}(1+2\a). \label{d1}
\ee
The particular values

\begin{eqnarray}
n=-\frac{4}{1+2\a}\,\,\,{\rm and}\,\,\,n=\frac{2}{1+2\a} \label{d2}
\end{eqnarray}
 give $\bar \Delta_-=0, ~\Delta_+=3$ and $\bar \Delta_+=0, ~\Delta_-=3$, respectively.
 Due to a coupling to a time-dependent function, the  helicities $\pm1$ of the massive spin-1 perturbation are constant on scales larger than the Hubble radius. It might be surprising that a constant super-Hubble mode is found for any value of $\alpha$. However, this is just 
 a consequence of the fact that $\alpha$ parametrizes the arbitrary mass (\ref{M12}) of the photon and the two possible values of $n$ become

 \begin{eqnarray}
n=1\pm \sqrt{4 M_1^2/H^2-3}.
\end{eqnarray}
The solutions to Eqs. (\ref{pp1}) and (\ref{pp2}) with Bunch-Davies initial conditions  are easily found to be

\begin{eqnarray}
&&\sigma^0_{0,1}=A_0N_0 (-k\t)^{(3-n(1+\a))/2} H^{(1)}_{(1+n+n \a))/2}(-k\t),
 \label{s1s}\\
 &&\sigma^{\pm 1}_{1,1}=A_1N_1 (-k\t)^{(1-n)/2} H^{(1)}_{(1+n+2n \a))/2}(-k\t),
 \label{s1s}
\end{eqnarray}
where $A_{|\lambda|}=\exp(i\pi/2(1+n(1+\a(1+|\lambda|)/2))$. 
The coefficients $N_0$ and $N_1$ can be calculated after normalization of the solution. For this, we need an inner product, which can be defined once  a conserved current is specified. It is straightforward to verify that the current 
\begin{eqnarray}
J^\mu=e^{I}\left(h^\rho\nabla^\mu {\sigma_{\rho}^*}'-\s_\rho^*\nabla^\mu {h^\rho}'
+\a h^\mu \s_\rho^* \nabla^\rho I-\a{\sigma^\mu}^*h_\rho\nabla^\rho I\right),
\end{eqnarray}
is conserved on shell,  $\nabla_\mu J^\mu=0$. 
Then, we may define the inner product of $f_\mu,h_\mu$ as
\begin{eqnarray}
\Big< f_\mu|h_\nu\Big>=(-i)\int d\Sigma \sqrt{\hat{g}}\,n_\mu 
e^{I}\left(h^\rho\nabla^\mu {\sigma_\rho^*}'-\s_\rho^*\nabla^\mu {h^\rho}'
+\a h^\mu \s_\rho^* \nabla^\rho I-\a{\sigma^\mu}^*h_\rho\nabla^\rho I\right),
\end{eqnarray}
where $\Sigma$ is a spacelike hypersurface with normal $n^\mu$ and $\hat{g}$ is its induced metric.
Normalizing the solutions as 
\begin{eqnarray}
\Big<\s^{(\lambda)}_\mu(\t,\vec{x})\Big|\s^{(\lambda')}_{\nu}(\t,\vec{x}')\Big>
=\delta^{\lambda\lambda'}\delta^{(3)}(\vec{x}-\vec{x}'), 
\end{eqnarray}
we find that 

\begin{eqnarray}
\Big<\s^{(0)}_\mu(\t,\vec{k})e^{i \vec{k}\cdot \vec{x}}\Big|\s^{(0)}_{\nu}(\t,\vec{k}')e^{i \vec{k}'\cdot \vec{x}}\Big>=0,\nonumber\\
\Big<\s^{(\pm 1)}_\mu(\t,\vec{k})e^{i \vec{k}\cdot \vec{x}}\Big|\s^{(\pm 1)}_{\nu}(\t,\vec{k}')e^{i \vec{k}'\cdot \vec{x}}\Big>=\frac{4k}{\pi}N_1^2 \delta^{(3)}(\vec{k}-\vec{k}').
\end{eqnarray}
The fact that the helicity-0 mode $\s^{(0)}_\mu(\t,\vec{k})$ has zero norm signals  a gauge symmetry of the field equation (\ref{gen1}). Indeed, it be straightforward to verify that Eq. (\ref{gen1}) is invariant under the gauge transformation 

\begin{eqnarray}
\s_\mu\to \s_\mu+\partial_\mu \theta +(1+\a)\theta \nabla_\mu I.
\end{eqnarray}
Due to this symmetry,  only the  $\pm 1$ helicities are propagating as we found above. Note that for $\alpha=-1$, we recognize the standard $U(1)$ gauge transformation of the gauge potential.

\subsection{Long-lived spin-1 perturbations and enhanced symmetry}
The special case $\alpha=-1$ is particularly interesting. From  Eq. (\ref{i10}) we immediately read off that $\beta=0$ and Eq. (\ref{m0m}) simplifies to
\begin{eqnarray}
(3H^2-M_1^2)\nabla_\mu I-\nabla_\mu M_1^2=0. 
\end{eqnarray}
Once $I$ is given, we can solve it to find $M_1^2$. However, there is also a solution independent of $I$, which is simply 
\begin{eqnarray}
M_1^2=3H^2. 
\end{eqnarray}
The equation of motion further reduces to   
\begin{eqnarray}
\square\sigma^\rho+\left(\nabla^\mu I\right) \nabla_\mu \sigma^\rho-
\nabla_\mu I \nabla^\rho \sigma^\mu-3H^2\sigma^\rho=0  \label{eq0}
\end{eqnarray}
and it is  straightforward to check, making use of Eq. (\ref{com}),  that Eq. (\ref{eq0}) is invariant under the gauge transformation
\begin{eqnarray}
\delta \sigma^\mu=\partial^\mu \xi, 
\end{eqnarray}
and therefore, it propagates two degrees of freedom,  corresponding to a massless photon. 
For all other values of $\alpha$ and/or  $M_1^2$, the gauge invariance is lost and we have the usual three degrees of freedom of a massive photon. 
\vskip 0.2cm
\begin{framed}
{\footnotesize
\noindent
Let us compare Eq. (\ref{eq0}), which describes a massless photon, with  the equation of that of an abelian vector $\sigma_\mu$  non-minimally coupled to the classical value of the inflaton field  $\phi_0(\tau)$ \cite{v1,v2}
\begin{eqnarray}
S=-\frac{1}{4}\int \d^4 x\sqrt{-g} \,J(\phi)F_{\mu\nu}F^{\mu\nu},\,\,\,\,\, F_{\mu\nu}=\partial_\mu\sigma_\nu-\partial_\nu\sigma_\mu.
\end{eqnarray}
The equation of motion for $\sigma_\mu$ is 
\begin{eqnarray}
J\nabla_\mu F^{\mu\nu}+(\nabla_\mu J)F^{\mu\nu}=0,
\end{eqnarray}
or
\begin{eqnarray}
\square \sigma_\nu-\nabla_\mu \nabla^\nu  \sigma^\mu+(\nabla_\mu I)\nabla^\mu  \sigma^\nu
-(\nabla_\mu I)\nabla^\nu  \sigma^\mu=0,  \label{eq1}
\end{eqnarray}
where $I=\ln J$. In the $\nabla_\mu  \sigma^\mu=0$ gauge
Eq. (\ref{eq1}) is written as 
\begin{eqnarray}
\square \sigma^\nu+(\nabla_\mu I)\nabla^\mu \sigma^\nu
-(\nabla_\mu I)\nabla^\nu \sigma^\mu-3H^2 \sigma^\nu=0, ~~~\nabla_\mu \sigma^\mu=0 \label{eq2}
\end{eqnarray}
which is identical to Eq. (\ref{eq0}). }
\end{framed}
\noindent
Eq. (\ref{eq0}) together the constraint are explicitly written as 

\begin{eqnarray}
\sigma_i''-\nabla^2\sigma_i-2\nabla_i \partial_i\sigma_\t+I'(\sigma_i'-\partial_i \sigma_\t)&=&0,\, \\
\sigma_\t''-\nabla^2\sigma_\t-\frac{2}{\t}\partial_i\sigma_i-\frac{2}{\t^2}\sigma_\t&=&0, \\
\sigma_\t'-\frac{2}{\t}\sigma_\t&=&\partial_i\sigma_i. 
\end{eqnarray}
Expanding in helicity modes, we find
\begin{eqnarray}
{\sigma^{\pm 1}_{1,1}}''+k^2\sigma^{\pm 1}_{1,1}+I'{\sigma^{\pm 1}_{1,1}}'&=&0, \label{s1p}\\
{\sigma^{0}_{0,1}}''-\frac{2}{\t}{\sigma^{0}_{0,1}}'+\left(k^2+\frac{2}{\t^2}\right)\sigma^{0}_{0,1}&=&0. \label{s2p}
\end{eqnarray}
Using the function $I$ of Eq. (\ref{I}), we find that  
\begin{eqnarray}
\Delta_-= 2-n ~~~{\rm and} ~~~ \Delta_+=1,
\label{pp}
\end{eqnarray}
in agreement with Eq. (\ref{cd}) once $\alpha=-1$ is taken.
%
%
Then, the  canonically normalized field $\bar{\sigma}_i$  (\ref{nor}) has 
\be
\bar\Delta=2-\frac{n}{2}.
\ee
If we wish a constant magnetic field 

\be
B_i={\rm exp}(I(\phi_0)/2)\,\epsilon_{ijk}\frac{\partial_j A_k}{a^2}\sim \tau^{n/2+2+\Delta-1}, 
\ee
we find two possible solutions

\be
B_i\sim \tau^{n/2+2+\Delta_{\pm} -1}=\left\{
\begin{array}{l}
\tau^{n/2+2}\Rightarrow n=-4,\\
\tau^{-n/2+3}\Rightarrow n=6
\end{array}, \right.
\ee
In the first case, however, a too large electromagnetic coupling constant is generated during inflation, while the second case implies a too large energy density in the electric modes. If we wish the  electric field 

\be
E_i={\rm exp}(I(\phi_0)/2)\,\frac{A'_i}{a^2}\sim \tau^{n/2+\Delta},
\ee
to be constant on super-Hubble scales, this implies
which implies

\be
E_i\sim \tau^{n/2+\Delta_{\pm}}=\left\{
\begin{array}{l}
\tau^{n/2+1}\Rightarrow n=-2,\\
\tau^{-n/2+2}\Rightarrow n=4
\end{array}, \right.
\ee
We recover the very well-know result that a massless photon coupled to the
inflaton field  in a proper way has super-Hubble perturbations which remain frozen during inflation \cite{v1,v2}.

\section{The spin-2 case}
Motivated by our findings for the spin-1 case, we now proceed to  consider a spin-2 field $\sigma^{\rho\sigma}$ with mass $m$ on the de Sitter background, with again the goal of investigating if it possible to couple it suitably to a function of the inflaton field in such a way that its canonically normalized helicities $\pm 2$ super-Hubble perturbations can stay constant in time. The equation and the constraints read

\begin{eqnarray}
\left(\square-m_2^2\right)\sigma^{\rho\sigma}=0, ~~~\nabla_\mu \sigma^{\mu\rho}=0,
~\sigma={\sigma^\rho}_\rho=0, 
\end{eqnarray}
where 
\begin{eqnarray}
m_2^2=m^2+2H^2. 
\end{eqnarray}
Let us now write,  as we did in Section 1 for the spin-1 case, the most  general coupling of the spin-2 field
$\sigma^{\rho\sigma}$ to functions of the  inflaton field $\phi_0(\tau)$

\begin{eqnarray}
&&\square\sigma^{\rho\sigma}+\left(\nabla^\mu I\right) \nabla_\mu 
\sigma^{\rho\sigma}+\alpha \left(\nabla_\mu I\right) \nabla^\rho\sigma^{\mu\sigma}+\alpha
\left(\nabla_\mu I\right)\nabla^\sigma\sigma^{\mu\rho}-M_2^2\sigma^{\rho\sigma}=0,
\label{s0}
\end{eqnarray}
where $\alpha$ is a numerical  constant. 
%
Taking the divergence of Eq.  (\ref{s0}), we find that 

\begin{eqnarray}
0&=&\nabla^\mu\sigma^{\rho\sigma}\Big[(1+\alpha)(\nabla_\rho\nabla_\mu I-\alpha \nabla_\mu I \nabla_\rho I)
\Big]+\alpha \nabla^\sigma\sigma^{\mu\rho}\left(\nabla_\rho\nabla_\mu I-\alpha \nabla_\mu I\nabla_\rho I\right)\nonumber \\
&+&\sigma^{\mu\sigma}\Big[(4H^2(1+\alpha)+\alpha M_2^2)\nabla_\mu I
-\nabla _\mu M_2^2
\Big].
\label{s000}
\end{eqnarray}
In order  not to  introduce any extra constraint on $\sigma^{\mu\nu}$ we should demand that 
\begin{eqnarray}
&&\nabla_\mu\nabla_\rho I-\alpha \nabla_\mu I \nabla_\rho I=I_0\, g_{\mu\rho}, \label{s1}\\
&&(4H^2(1+\alpha)+\alpha M_2^2)\nabla_\mu I
-\nabla _\mu M_2^2=0. \label{m0m2}
\end{eqnarray}
%
The function $I_0$ can be directly obtained by choosing $\mu=\rho=i$ and the  condition (\ref{s1}) is satisfied if 
\begin{eqnarray}
I''+\frac{2}{\tau}I'-\alpha {I'}^2=0, 
\end{eqnarray}
which gives 
\begin{eqnarray}
I=A-\frac{1}{\alpha}\ln\left(B+\frac{\alpha}{H \tau}\right), \label{i2}
\end{eqnarray}
where $A$ and $B$ are numerical constants. 
 Eq. (\ref{m0m2}) specifies  $M_2^2$ to be 
\begin{eqnarray}
  M_2^2=-\frac{4(1+\alpha)}{\alpha}H^2, ~~~\mbox{or}~~~M_2^2=\frac{m^2\tau
  -4(1+\alpha)H^2}{\alpha+A H \tau} 
  \end{eqnarray}  
and  the equations of motion (\ref{s0}) are
\begin{eqnarray}
&&\sigma_{\t\t}''+\frac{2}{\tau}\sigma_{\t\t}'-\left(\partial_i^2-\frac{M_2^2/H^2-8}{\tau^2}\right)\sigma_{\t\t}\nonumber \\
&&+(1+2\alpha)I'\left(\sigma_{\t\t}'+\frac{2}{\t}\s_{\t\t}\right)=\frac{4}{\tau}\partial_i \sigma_{0i}+\frac{2}{\tau^2}\sigma_{ii},\\
&&\s_{\t i}''+\frac{2}{\tau}\s_{\t i}'-\left(\partial_i^2-\frac{{M_2}^2/H^2-8}{\tau^2}\right)\sigma_{\t i}\nonumber \\
&&+I'\left\{(1+\a)\s_{\t i}'+\frac{2(1+2\a)}{\t}\s_{\t i}+\a \partial_i \s_{\t\t}\right)=\frac{2}{\t} \del_i\s_{\t\t}+\frac{2}{\tau}\del_j\s_{ij},\\
&&\s_{ij}''+\frac{2}{\t}\s_{ij}'-\left(\partial_i^2-\frac{{M_2}^2/H^2-4}{\tau^2}\right)\sigma_{ij}\nonumber\\
&&+I'\left(\s_{ij}'+\a(\partial_i \s_{\t j}+\partial_j\s_{\t i})+\frac{2(1+\a)}{\tau}\s_{ij}+\frac{2\a}{\tau}\delta_{ij}\s_{\t\t}\right)=\frac{4}{\t}\partial_{(i}\s_{j)0}+\frac{2}{\t^2}\delta_{ij}\s_{\t\t}.
\end{eqnarray}
In addition, the helicity fields $\sigma_{\t\t}$, $\sigma_{0i}$ and $\sigma_{ij}$ are subject to the constraints,  which are  explicitly written as
\begin{eqnarray}
\sigma_{\t\t}'-\del_i\s_{\t i}-\frac{1}{\t}\left(\s_{\t\t}+\s_{ii}\right)&=&0,\label{c11}\\
\s_{\t i}'-\del_j\s_{ij}-\frac{2}{\t}\s_{\t i}&=&0, \label{c22} \\
\s_{\t\t}-\s_{ii}&=&0.  \label{c33}
\end{eqnarray}
We may now expand  the Fourier modes of $\sigma_{\mu\nu}$ in helicity eigenstates as 
\begin{eqnarray}
\sigma_{\mu\nu}=\sum_{\lambda=-2}^2\sigma^{(\lambda)}_{\mu\nu}. 
\end{eqnarray}
The mode functions can then be written in terms of the various helicities as
($\widehat{\sigma}_{ij}=\sigma_{ij}\!-\!\delta_{ij}\sigma_{\t\t}/3$) \cite{baumann}
\begin{eqnarray}
~~~~~\sigma_{\t\t}^{(0)}=\sigma^0_{0,2},~\,~&\sigma_{\t\t}^{(\pm 1)}=0,~~~~&\sigma_{\t\t}^{(\pm 2)}=0,\\~~
\sigma_{i\t}^{(0)}=\sigma^0_{1,2}\varepsilon^0_i,\,&\sigma_{i\t}^{(\pm 1)}=\sigma^{\pm1}_{1,2}\varepsilon^{\pm 1}_i,
&\sigma_{i\t}^{(\pm 2)}=0,\\
\widehat{\sigma}_{ij}^{(0)}=\sigma^0_{0,2}\varepsilon^0_{ij},&\widehat{\sigma}_{ij}^{(\pm 1)}=\sigma^{\pm1}_{2,2}\varepsilon^{\pm 1}_{ij},
&\widehat{\sigma}_{ij}^{(\pm 2)}=\sigma^{\pm2}_{2,2}\varepsilon^{\pm 2}_{ij},
\end{eqnarray}
where the polarizations tensors are given by
\begin{eqnarray}
\varepsilon^0_i=\hat{k}_i, ~~~ 
\varepsilon^0_{ij}=\frac{3}{2}\left(\hat{k}_i\hat{k}_j-\frac{1}{3}\delta_{ij}\right), ~~~\varepsilon^{\pm 1}_{ij}=\frac{3}{2}
\left(\hat{k}_i\varepsilon^{\pm 1}_j+\hat{k}_j\varepsilon^{\pm 1}_i\right),
\end{eqnarray}
and $\varepsilon^{\pm 1}_{i}, \varepsilon^{\pm 2}_{ij}$ are such that 
\begin{eqnarray}
\hat{k}_i\,\varepsilon^{\pm 1}_i =0, ~~~\hat{k}_i\varepsilon^{\pm 2}_{ij}=0,~~~ \varepsilon^{\pm 2 *}_{ij}=
\varepsilon^{\mp 2}_{ij}, ~~~\varepsilon^{\pm 2}_{ij}\varepsilon^{\mp 2}_{ij}=4. 
\end{eqnarray}
In addition, they satisfy
\begin{eqnarray}
\hat{k}_i \varepsilon^0_{ij}=\varepsilon^0_j,~~~\hat{k}_i\varepsilon^{\pm 1}_{ij}=\frac{3}{2}\varepsilon^{\pm 1}_j,~~~\hat{k}_i\varepsilon^{\pm 2}_{ij}=0, 
\end{eqnarray}
and for momentum along the $z$-axis we may take 
\begin{eqnarray}
 \varepsilon^{\pm 1}_i=(1,\pm i,0), ~~~\varepsilon^{\pm 2}_{ij}=
 \left(\begin{array}{ccc}
1&\pm i&0\\
 \pm i&-1&0\\
 0&0&0
 \end{array}\right).
 \end{eqnarray} 
 The equations for the different helicity modes become
\begin{eqnarray}
{\s_{2,2}^{\pm 2}}''+\left(\frac{2}{\tau}+I'\right){\s_{2,2}^{\pm 2}}'+\left(k^2+\frac{{M_2}^2/H^2-4}{\tau^2}+\frac{2(1+\a)}{\t}I'\right){\s_{2,2}^{\pm 2}}&=&0,\label{h1}\\
{\s_{1,2}^{\pm 1}}''+(1+\a)I'{\s_{1,2}^{\pm 1}}'+\left(k^2+\frac{{M_2}^2/H^2-4}{\tau^2}+\frac{2(1+2\a)}{\t}I'\right){\s_{1,2}^{\pm 1}}&=&0, \label{h2}
\\
{\s_{0,2}^{ 0}}''-\left(\frac{2}{\t}-(1+2\a)I'\right){\s_{0,2}^{\pm 0}}'+\left(k^2+\frac{{M_2}^2/H^2-2}{\tau^2}+\frac{2(1+2\a)}{\t}I'\right){\s_{0,2}^{0}}&=&0. \label{h3}
\end{eqnarray}
They admit scaling solutions  when $M_2^2$ is constant 
\begin{eqnarray}
M_2^2=\frac{1}{\alpha}(s+2)(\alpha-s\alpha-1)H^2=-\frac{4(1+\a)}{\a}H^2, \label{mms1}
\end{eqnarray}
and $I$ has the form 
\begin{eqnarray}
I=A+\frac{1}{\alpha}\ln(-H \tau)=A+\frac{1}{\a}\int^\phi \frac{V(\phi')}{V'(\phi')}\d \phi'. \label{II0}
\end{eqnarray}
Then, Eqs. (\ref{h1}), (\ref{h2}),  and Eq. (\ref{h3}) reduce to   
\begin{eqnarray}
{\s_{2,2}^{\pm 2}}''+\left(\frac{2+1/\a}{\tau}\right){\s_{2,2}^{\pm 2}}'+\left(k^2+\frac{{M_2}^2/H^2-4+2(1+\a)/\a}{\tau^2}\right){\s_{2,2}^{\pm 2}}&=&0,\label{h11}\\
{\s_{1,2}^{\pm 1}}''+\frac{1+1/\a}{\t}{\s_{1,2}^{\pm 1}}'+\left(k^2+\frac{{M_2}^2/H^2-4+2(1+2\a)/\a}{\tau^2}\right){\s_{1,2}^{\pm 1}}&=&0, \label{h22}
\\
{\s_{0,2}^{ 0}}''-\frac{2-(1+2\a)/\a}{\t}{\s_{0,2}^{\pm 0}}'+\left(k^2+\frac{{M_2}^2/H^2-2+2(1+2\a)/\a}{\tau^2}\right){\s_{0,2}^{0}}&=&0.  \label{h33}
\end{eqnarray}
In addition, the conditions (\ref{c11}), (\ref{c22}), and (\ref{c33}) turn out to be  
\begin{eqnarray}
{\s_{1,2}^{ 0}}&=&-\frac{i}{k}\left({\s_{0,2}^{ 0}}'-
\frac{2}{\t}{\s_{0,2}^{ 0}}\right),\nonumber \\
{\s_{2,2}^{ 0}}&=&-\frac{i}{k}\left({\s_{1,2}^{ 0}}'-
\frac{2}{\t}{\s_{1,2}^{ 0}}\right)-\frac{1}{3}{\s_{0,2}^{ 0}},\nonumber \\
{\s_{2,2}^{ \pm 1}}&=&-\frac{i}{k}\left({\s_{1,2}^{ \pm1}}'-
\frac{2}{\t}{\s_{1,2}^{ \pm 1}}\right)
\end{eqnarray}
with  solutions 
\begin{eqnarray}
\label{gh}
{\s_{\lambda,2}^{\pm \lambda}}&=&N_2^{\pm \lambda} 
(-k\tau)^{\frac{\a(1-\lambda)-1}{2\a}}
H_{\nu_{\lambda,2}}^{(1)}(-k\t),\,\,\,\, \nu_{\lambda,2}=
\left|\frac{1+(3+\lambda)\a}{2\a}\right|.
\end{eqnarray}
\begin{framed}
{\footnotesize
\noindent
The case  $\alpha=-1$  can indeed be obtained from the 
standard massive spin-2 action by multiplying the latter by the factor $I$. 
Indeed, let us consider the action

\begin{eqnarray}
 S=\frac{1}{2}\int \d^4 x\sqrt{g}\, \left\{
{\sigma}_{\mu\nu}\widehat{{\cal E}}^{\mu\nu\rho\sigma}{\sigma}_{\rho\sigma}-\frac{m^2}{2} J\left({\sigma}^{\mu\nu}\s_{\mu\nu}
-{\sigma}^2\right)\right\},\label{m2}
\end{eqnarray}
where  $I=\ln J$, $\sigma={\sigma^\mu}_\mu$ and 
\begin{eqnarray}
\widehat{{\cal E}}^{\mu\nu\rho\sigma}{\sigma}_{\rho\sigma}&=&-\frac{1}{2}\nabla_\lambda\big{(}
J\nabla^\mu {\sigma}^{\nu\lambda}\br-\frac{1}{2}\nabla_\lambda\big{(}
J\nabla^\nu {\sigma}^{\mu\lambda}\br+\frac{1}{2}
\nabla_\lambda\bl J\nabla^\lambda {\sigma}^{\mu\nu}\br+\frac{1}{2}\nabla^\mu\bl J\nabla^\nu {\sigma}\br\nonumber \\
&&-\frac{1}{2}g^{\mu\nu}\Bl\nabla_\lambda\bl J\nabla^\lambda {\sigma}\br-\nabla_\lambda\bl J\nabla_\rho {\sigma}^{\lambda\rho}\br\Br+3H^2 J
\left(\sigma^{\mu\nu}-g^{\mu\nu }\sigma\right). \label{em2}
\end{eqnarray}
The kinetic part of the action (\ref{m2})  can in fact be written as the quadratic part of the Einstein-Hilbert action in the Jordan frame

\begin{eqnarray}
\frac{1}{2}\sqrt{-\hat{g}}\, J\, (\hat{R}-2\Lambda)=\sqrt{-g}\left(
\frac{J}{2}(R-2\Lambda)-\frac{J}{2}(G^{\mu\nu}+\Lambda g^{\mu\nu})\sigma_{\mu\nu}+\frac{1}{2}
{\sigma}_{\mu\nu}\widehat{{\cal E}}^{\mu\nu\rho\sigma}{\sigma}_{\rho\sigma}+\cdots\right),
\end{eqnarray}
 where $\hat g _{\mu\nu}=g_{\mu\nu}+\sigma_{\mu\nu}$ and $g_{\mu\nu}$ is the background de Sitter metric. It is straightforward to check by taking the 
 divergence, the double divergence and the trace of Eq.(\ref{em2}) that we still get the conditions
 \begin{eqnarray}
 \nabla_\mu\sigma^{\mu\nu}=0, ~~~\sigma=0. 
 \end{eqnarray}
Using the above constraints, it is easy to verify that the equations of motion reduces to Eq.(\ref{s0}) for $m^2=M_2^2-2H^2=-2H^2$ (for $\a=-1$). 
One may also calculate the corresponding energy-momentum tensor for $\sigma_{\mu\nu}$ by varying the action (\ref{m2}) with respect to the background metric. The result is 

\begin{eqnarray}
T_{\mu\nu}&=&J \nabla_\lambda \sigma_{\mu\kappa}\nabla^\lambda{\s_\nu}^\kappa
+J\nabla_\mu \s_{\kappa\lambda}\nabla_\nu \s^{\kappa\lambda}-2J\nabla^\kappa{\s_{(\mu}}^\lambda\nabla_{\nu)}\s_{\kappa\lambda}-J\nabla_\lambda\s_{\kappa(\mu}\nabla^\kappa{\sigma_{\nu)}}^\lambda
\nonumber \\
&&-8H^2J\left(\s_{\mu\kappa}{\s_\nu}^\kappa-\sigma \s_{\mu\nu}\right)
+\frac{1}{4}g_{\mu\nu}J\left[\nabla_\lambda \s_{\s\rho}\nabla^\sigma\s^{\rho\lambda}+\nabla_\lambda \s_{\sigma\rho}\nabla^\rho\s^{\s\lambda}\right.\nonumber \\
&&\left.-\nabla_\lambda\s_{\s\rho}\nabla^\lambda\s^{\s\rho}+4H^2(\s_{\kappa\lambda}\s^{\kappa\lambda}-\s^2)
\right].
\end{eqnarray}}

\end{framed}

\subsection{Orthonormality of the mode functions and inner product}
The various coefficients $N_2$ are to  be specified  by the requirement of orthonormality of the mode functions

\begin{eqnarray}
\Big< \sigma_{\mu\nu}^{(\lambda)}(\vec{k},\t)e^{i\vec{k}\cdot \vec{x}}|
\sigma_{\rho\sigma}^{(\lambda')}(\vec{k},\t)e^{i\vec{k'}\cdot \vec{x}}\Big>=\delta_{\lambda\lambda'}\delta^{(3)}(\vec{k}-\vec{k}').
\end{eqnarray}
Therefore, we need to define first the inner product. The latter can be defined once a conserved current is found. It can be check that 
 the current 
\begin{eqnarray}
J^\mu&=&e^{I}\Big[h^{\rho\sigma}\nabla^\mu \sigma^*_{\rho\sigma}-
\sigma^*_{\rho\sigma}\nabla^\mu h^{\rho\sigma}+
2\a \nabla_\sigma I\Big{(}{{\sigma^*}_\rho}^\sigma h^{\mu\rho}-{\sigma^*_\rho}^\mu h^{\rho\sigma}\Big{)}\Big],\nonumber\\
&&
\end{eqnarray}
is conserved so that $\nabla_\mu J^\mu=0$. Therefore, we may define the inner product as

\begin{eqnarray}
\Big< h^{\mu\nu}|f_{\rho\sigma}
\Big>&=&
(-i)\int \d\Sigma\sqrt{\hat{g}} \, n_\mu 
 \, e^{I}\Big{\{}\big{(}h^{\rho\sigma}\nabla^\mu f^*_{\rho\sigma}-
f^*_{\rho\sigma}\nabla^\mu h^{\rho\sigma}\Big{)}
+2\a \nabla_\sigma I\Big{(}{{f^*}_\rho}^\sigma 
h^{\mu\rho}-{f^*_\rho}^\mu h^{\rho\sigma}\Big{)}\Big{\}},\nonumber\\
&&\label{inner}
\end{eqnarray}
where $\Sigma$ is a spacelike hypersurface with normal vector $n^\mu$ and 
$\hat{g}$ is the determinant of the induced metric on the hypersurface. 
Since $I$ is a function of time only, this implies that 

\begin{eqnarray}
\Big< \sigma_{\mu\nu}^{(\lambda)}(\vec{k},\t)e^{i\vec{k}\cdot \vec{x}}\Big|
\sigma_{\rho\sigma}^{(\lambda')}(\vec{k},\t)e^{i\vec{k'}
\cdot \vec{x}}\Big>&=&
(-i) (H\t)^2 \eta^{\mu\rho}\eta^{\nu\sigma}\nonumber \\
&&\int \d ^3 x(-H \t)^{1/\alpha}\left(\sigma_{\mu\nu}^{(\lambda)}{{\sigma^*}_{\rho\sigma}^{(\lambda)}}'-
{\sigma^*}_{\rho\sigma}^{(\lambda)}{{\sigma}_{\mu\nu}^{(\lambda)}}'\right)e^{i(\vec{k}-\vec{k})'\cdot\vec{x}}\nonumber
\end{eqnarray}
and 

\begin{eqnarray}
\Big< \sigma_{\mu\nu}^{(0)}(\vec{k},\t)e^{i\vec{k}\cdot \vec{x}}\Big|
\sigma_{\rho\sigma}^{(0)}(\vec{k},\t)e^{i\vec{k'}\cdot \vec{x}}\Big>&=&
\frac{2(1+6\a+8\a^2)}{\a^2\pi}\frac{H^2}{k}\left(\frac{H}{k}\right)
^{1/\a}{(N_2^0)}^2\delta^{(3)}(\vec{k}-\vec{k}'),\nonumber\\
&&\\
\Big< \sigma_{\mu\nu}^{(\pm 1)}(\vec{k},\t)e^{i\vec{k}\cdot \vec{x}}\Big|
\sigma_{\rho\sigma}^{(\pm 1)}(\vec{k},\t)e^{i\vec{k'}\cdot \vec{x}}\Big>&=&
0, \\
\Big< \sigma_{\mu\nu}^{(\pm 2)}(\vec{k},\t)e^{i\vec{k}\cdot \vec{x}}\Big|
\sigma_{\rho\sigma}^{(\pm 2)}(\vec{k},\t)e^{i\vec{k'}\cdot \vec{x}}\Big>&=&
\frac{4H^2}{k\pi}\left(\frac{H}{k}\right)^{1/\a}
{(N_2^{\pm 2})}^2\delta^{(3)}(\vec{k}-\vec{k}'),
\end{eqnarray}
from where we find 
\begin{eqnarray}
\label{ghh}
N_2^0&=&\sqrt{\frac{\pi}{2H}}\frac{\a}{\sqrt{(1+6\a+8\a^2)}}\left(\frac{k}{H}\right)^{\frac{\a+1}{2\a}}, \\
N_2^{\pm 2}&=&\sqrt{\frac{\pi}{4H}}\left(\frac{k}{H}\right)^{\frac{\a+1}{2\a}}.
\end{eqnarray}
\begin{framed}
{\footnotesize
\noindent
 The fact that the norm of the helicity $\pm 1$ modes vanish for any value of $\alpha$ means that there should also exist a gauge invariance  projecting out the $\pm 1$ helicity modes. It is straightforward to find that 
for 
\begin{eqnarray}
\delta\sigma_{\mu\nu}= -
\xi_\mu\nabla_\nu J-\xi_\nu \nabla_\mu J,~~~J=e^{\a I},  \label{gauge2}
\end{eqnarray}
the following equation is satisfied
\begin{eqnarray}
0&=&\square\delta\sigma_{\rho\sigma}+\left(\nabla^\mu I\right) \nabla_\mu 
\delta\sigma_{\rho\sigma}+\alpha \left(\nabla^\mu I\right) \nabla^\rho\delta\sigma_{\mu\sigma}+\alpha
\left(\nabla^\mu I\right)\nabla_\sigma\sigma_{\mu\rho}-M_2^2\delta\sigma_{\rho\sigma}\nonumber \\
&=&-\frac{(1+2\a)}{\a}\Big{(}\nabla_\rho J\nabla_\sigma J\nabla_\mu\xi^\mu+\sigma\leftrightarrow \rho\Big{)}. 
\end{eqnarray}
Therefore, the transformation (\ref{gauge2}) is a gauge transformation if it satisfies 
\begin{eqnarray}
\xi^\t=0~~~{\rm and}~~~\nabla_i\xi^i=0,
\end{eqnarray}
where the vanishing of the temporal component of $\xi^\mu$ follows again from the traceleness condition $\delta\sigma^{\mu}_{\mu}=-2\nabla_\mu J\xi^\mu=0$. Thus, the gauge parameter satisfies two conditions leading to $4-2=2$  free gauge parameters, which leads to $5-2=3$ polarizations, precisely the  helicity 0 and $\pm 2$ ones. 
}
\end{framed}
\noindent
\subsection{The Higuchi bound and the long-lived spin-2 perturbations}
The two-point function for the Fourier modes of the spin-$s$ field 
$\sigma_{\mu_1\cdots \mu_s}(\vec{k},\tau)$ can be expressed in terms of a null polarization vector $\varepsilon_i$ $(\varepsilon_i^2=0)$. For momentum $\vec{k}=(0,0,k)$ along the $z$-axis,  we may choose $\varepsilon_i=(\cos\psi,\sin\psi,i)$, 
$\widetilde\varepsilon_i=(\cos\psi',\sin\psi',-i)$  and the two-point function can be expressed as 
\begin{eqnarray}
\Big< \varepsilon^s\sigma^s(\tau)\, \widetilde{\varepsilon}^s\sigma^s(\tau')\Big>'=\sum_{\lambda=-s}^{s}e^{i\lambda(\psi-\psi')}
\left(\frac{(2s-1)!!}{(2\lambda-1)!!(s-\lambda)!}\right)^2\sigma^\lambda_{s,s}
(-k\tau){\sigma^*}^{\lambda}_{s,s}(-k\tau'), 
\end{eqnarray}
where $\varepsilon^s\sigma^s(\tau)=\varepsilon^{i_1}\cdots \varepsilon^{i_s}
\sigma_{i_1\cdots i_s}(\vec{k},\tau)$. 
For the $s=2$ under consideration, positivity of the two-point function then leads to the positivity of the squares of $N_2^0$ and $N_2^{\pm 2}$. In particular, the positivity of the square of $N_2^0$ leads to the condition
\begin{eqnarray}
•1+6\alpha+8\alpha^2>0, 
\end{eqnarray}
which is satisfied for $\alpha$ in the range
\begin{eqnarray}
\a<-\frac{1}{2} ~~~\mbox{or}~~~\a>-\frac{1}{4}. \label{bound}
\end{eqnarray}
This is the corresponding Higuchi bound  for  spin-2 fields coupled non-trivially to the inflation field. The next step is to calculate the 
 scaling dimension of the spin-2 fields. We look for solutions of the form 
\begin{eqnarray}
\sigma_{ij}(\vec{x},\tau)=\sigma^+_{ij}(\vec{x})\tau^{\Delta_+-2}
+\sigma^-_{ij}(\vec{x})\tau^{\Delta_--2}. 
\end{eqnarray}
Then we find that 
\begin{eqnarray}
\Delta_-=-1-\frac{1}{\a}, ~~~\Delta_+=4.
\end{eqnarray}
Going to canonically normalized fields $\bar{\sigma}_{ij}$, it  is easy to convince oneself that 

\be
\bar{\sigma}_{ij}={\rm exp}(I(\phi_0)/2){\sigma}_{ij}.
\ee
Indeed, from Eq. (\ref{s0}) one can see that with this choice the first two terms combine to give only $\square \bar{\sigma}_{ij}$ (plus other interaction pieces).
Being the scaling dimension of $\bar{\sigma}_{ij}$ equal to $\bar\Delta-2$, we have two options. The first one is 

\be
\bar\Delta_--2=\Delta_- -2+\frac{1}{2\a}=-3-\frac{1}{2\alpha}.
\ee
Demanding $\bar\Delta_-=0$ to have long-lived perturbations, we get

\be
\alpha=-\frac{1}{2},
\label{vv}
\ee
corresponding to $\bar{\Delta}_+=3$ and which saturates the Higuchi bound (\ref{bound}). In fact this border limit introduces an extra symmetry, as we will discuss in the next subsection.
The second case is

\be
\bar\Delta_+ -2=\Delta_+ -2+\frac{1}{2\a}=4+\frac{1}{2\alpha}.
\ee
Demanding $\bar\Delta_+=0$ to get long-lived perturbations, we get

\be
\alpha=-\frac{1}{8},
\label{vv1}
\ee
corresponding to $\bar{\Delta}_-=3$ and which is also allowed by the Higuchi bound.

\subsection{Long-lived spin-2 perturbations and enhanced symmetry}
It is easy to see that for the value (\ref{vv})
also the norm of the zero-helicity state vanish, or equivalently, $N_2^0$ blows up. Therefore, for $\alpha=-1/2$, only the  $\pm 2$ helicities survive. In this case, there should be an further gauge symmetry. Indeed, for 

\begin{eqnarray}
\delta\sigma_{\mu\nu}= \nabla_\mu\xi_\nu+\nabla_\nu\xi_\mu-
\xi_\mu J^{-1}\nabla_\nu J-\xi_\nu J^{-1}\nabla_\mu J,~~~J=e^{\a I},  \label{gauge}
\end{eqnarray}
we find that 
\begin{eqnarray}
0&=&\square\delta\sigma_{\rho\sigma}+\left(\nabla^\mu I\right) \nabla_\mu 
\delta\sigma_{\rho\sigma}+\alpha \left(\nabla^\mu I\right) \nabla^\rho\delta\sigma_{\mu\sigma}+\alpha
\left(\nabla^\mu I\right)\nabla_\sigma\sigma_{\mu\rho}-{M}_2^2\delta\sigma_{\rho\sigma}\nonumber \\
&=&\frac{(1+2\a)}{J\a}\Big{(}\nabla_\mu J\, \nabla^\mu \nabla_\rho(J^{-1}\xi_\sigma)+3H^2J\nabla_\rho(J^{-1}\xi_\sigma)+\sigma\leftrightarrow \rho\Big{)}. 
\end{eqnarray}
Hence, for $\alpha=-1/2$, the equation for the spin-2 field (\ref{s0}) is invariant under the gauge transformation (\ref{gauge}).   As a result of the gauge invariance and the tracelessness condition 

\begin{eqnarray}
\nabla_\mu \xi^\mu=\xi^\mu \nabla_\mu J,
\end{eqnarray}
the gauge parameter provides $4-1=3$ free parameters which leads to 
$5-3=2$ propagating modes,  the helicities $\pm 2$. This result does not come as a surprise. Indeed, for  $\alpha=-1/2$ the linear equation for 
$\bar{\sigma}_{ij}=(-H\tau){\sigma}_{ij}$ reduces to the equation of motion for a massless graviton. In addition, the gauge symmetry (\ref{gauge}) is written 
 as 

\begin{eqnarray}
\delta\bar \sigma_{\mu\nu}= \nabla_\mu\epsilon_\nu+\nabla_\nu\epsilon_\mu, ~~~\epsilon_\mu=e^{I/2} \xi_\mu, \label{gauge1}
\end{eqnarray}
that is   the standard gauge transformation of a massless spin-2 field which,  for the cosmologically interesting case where the transformation is  done on fixed  space hypersurfaces,  leaves the helicity-2 field unchanged (at the linear level).
  So, by suitably coupling a massive spin-2 field to the inflation background one can obtain at the quadratic level an effectively massless helicity-2 state. Of course this degree of freedom
couples to the comoving curvature perturbation differently from the standard massless graviton.

\section{The spin-$s$ case}
Let us now consider a generic massive spin-$s$ field 
$\sigma^{\mu_1\cdots\mu_s}$ on a four-dimensional de Sitter spacetime. This field obeys 
 the equation of motion 
\begin{eqnarray}
\left(\square-m_s^2\right)\sigma^{\mu_1\cdots\mu_s}=0, 
\end{eqnarray}
where 
\begin{eqnarray}
m_s^2=m^2-(s^2-2s-2)H^2,
\end{eqnarray}
and the constraints 
\begin{eqnarray}
\nabla_{\mu_1}\sigma^{\mu_1\cdots\mu_s}={\sigma_{\mu_1}}^{\mu_1\mu_3\cdots\mu_s}=0,  \label{csp}
\end{eqnarray}
which ensure that there are $2s+1$ degrees of freedom. 
Again, we are interested to see if such field can coupled consistently to the inflaton field in such a way to obtain frozen perturbations on super-Hubble scales.
The general coupling  to the inflaton will have the form

\begin{eqnarray}
&&\square\sigma^{\mu_1\cdots\mu_s}+\left(\nabla^\mu I\right) \nabla_\mu 
\sigma^{\mu_1\cdots\mu_s}+\alpha \left(\nabla_\mu I\right) \nabla_{\mu_1} 
\sigma^{\mu\mu_2\cdots\mu_s} +\cdots \nonumber \\
&&\cdots +\alpha \left(\nabla_\mu I\right) \nabla_{\mu_s} 
\sigma^{\mu_1\cdots\mu_{s-1}\mu} \label{eqms}
-M_s^2(\phi)\sigma^{\mu_1\cdots\mu_s}=0, \label{ss}
\end{eqnarray}
where,   in order to have again the same degrees of freedom, we retain the constraints (\ref{csp}).
We have not included possible terms of the form $ I_{\mu_1\mu}\,\sigma^{\mu \mu_2\cdots\mu_s}+{\rm permutations}$, as the trace conditions  will demand $I_{\mu_1\mu}$ to be proportional to the metric (otherwise we have to  put extra conditions on the spin-$s$ field), and therefore such terms can be absorbed in the mass term. 
 
Taking the divergence of the equation of motion  (\ref{ss}), we get 

 \begin{eqnarray}
 0&=&\nabla_{\mu}\sigma^{\mu_1\cdots\mu_s}\Big{\{}(1+\alpha)(\nabla_\mu\nabla_{\mu_1} I-\alpha\nabla_\mu I\, \nabla_{\mu_1} I)\Big{\}}+
\nonumber \\
&&+ \beta \nabla^{\mu_1}\sigma^{\mu\mu_2\cdots\mu_{s}}\Big{(}(\nabla_\mu\nabla_{\mu_1} I-\alpha\nabla_\mu I\, \nabla_{\mu_1} I)\Big{)}\cdots\nonumber\\ 
&&\cdots + \beta \nabla^{\mu_s}\sigma^{\mu_1\cdots\mu_{s-1}\mu}\Big{(}(\nabla_\mu\nabla_{\mu_1} I-\alpha\nabla_\mu I\, \nabla_{\mu_1} I)\Big{)}+\nonumber\\
&&\Big{\{}\left[(s+2)H^2(1+s\alpha-\alpha)+\alpha M_2^2\right]\nabla_\mu I-\nabla_\mu M_s^2\Big{\}}\sigma^{\mu\mu_2\cdots\mu_s}.
\end{eqnarray}
%
This relation leads to 
\begin{eqnarray}
&&\nabla_\rho\nabla_\mu I-\alpha \nabla_\mu I \nabla_\rho I=I_0g_{\mu\nu}, \label{s101}\\
&&\left[(s+2)H^2(1+s\alpha-\alpha)+\alpha M_2^2\right]\nabla_\mu I-\nabla_\mu M_s^2=0. \label{m0}
\end{eqnarray}
Eq. (\ref{s101})  is satisfied if $I$ is given by Eq. (\ref{II0}), which we report here again
\begin{eqnarray}
I=A+\frac{1}{\alpha}\ln\left(-H \t\right), 
\end{eqnarray}
and the mass parameter turns out then to be
\begin{eqnarray}
&&
M_s^2=\frac{1}{\alpha}(s+2)H^2(\alpha-s\alpha-1)+e^{\alpha I}
m_0^2,
\end{eqnarray}
where $m_0^2$ is an integration constant. Note that  Eq. (\ref{m0})   admits the constant solution 
\begin{eqnarray}
M_s^2=\frac{1}{\alpha}(s+2)H^2(\alpha-s\alpha-1), \label{ms2}
\end{eqnarray}
corresponding to $m_0^2=0$. 
We may expand now $\sigma_{\mu_1\cdots \mu_s}$ in helicity modes as 
\begin{eqnarray}
\sigma_{\mu_1\cdots \mu_s}=\sum_{\lambda=-s}^s
\sigma^{(\lambda)}_{\mu_1\cdots \mu_s}.
\end{eqnarray}
In particular,  for a mode of helicity $\lambda$ and $n$-polarization directions we may write 
\begin{eqnarray}
\sigma^{(\lambda)}_{i_1i_2\cdots i_n\t\cdots\t}=\sigma^{(\lambda)}_{n,s}
\varepsilon{(\lambda)}_{i_1\cdots i_n}, ~~~\sigma^{(\lambda)}_{n,s}=0,~~n<|\lambda|, 
\end{eqnarray}
where $\varepsilon{(\lambda)}_{i_1\cdots i_n}$ are polarization tensors. 
Then from Eq. (\ref{ss}) we find that the  $n=|\lambda|$ helicity $\lambda$ mode function $\sigma^{(\lambda)}_{|\lambda|,s}$ satisfies the equation

\begin{eqnarray}
&&{\sigma^{(\lambda)}_{|\lambda|,s}}''-\left[\frac{2(1-\lambda)}{\t}-(1+\alpha(s-\lambda))I'\right]
{\sigma^{(\lambda)}_{|\lambda|,s}}'
\nonumber \\
&&+\left[k^2+\frac{M_s^2/H^2-s+\lambda(\lambda-3)}{\t^2}+\frac{s(1+\a s)+\a \lambda (1-\lambda)}{\t}I'\right]
\sigma^{(\lambda)}_{|\lambda|,s}=0.
\label{eqs}
\end{eqnarray}
We close this subsection with a  final comment concerning the possibility of superluminal propagation of the spin-$s$ field $\sigma_{\rho_1\cdots\rho_s}$. This is determined by the leading two derivative matrix $S^{\mu\nu}$ in the equation of motion 

\begin{eqnarray}
S^{\mu\nu}\partial_\mu \partial_\nu \sigma_{\rho_1\cdots\rho_s}+\cdots=0.
\end{eqnarray}
Since this  term is exactly the same with the leading two-derivative term when there is no  coupling to the inflaton, we conclude that the coupling of the spin-$s$ field to the inflaton does not change its superluminality properties. Therefore,  the spin-$s$ field $\sigma_{\rho_1\cdots\rho_s}$ propagates causally even when it is coupled to the inflaton field as long as  this coupling is of the form considered here.

\subsection{Long-lived spin-$s$ perturbations}
Having found the generic equation of motion for a spin-$s$ field, we are now ready to look for frozen super-Hubble modes.
Eq. (\ref{eqs}) admits scaling solutions only when $M_s^2$ is constant 
and it must therefore be  given by the expression (\ref{ms2}). Indeed, looking for solutions of the form 
\begin{eqnarray}
\sigma^{(\lambda)}_{|\lambda|,s}(\t,\vec x)=\tau^{\Delta-s} \widetilde\sigma^{(\lambda)}_{|\lambda|,s}(\vec x),
\end{eqnarray}
we find 
\begin{eqnarray}
\Delta_-=1-\lambda-\frac{1}{\a}, ~~~\Delta_+=2+s.
\end{eqnarray}
The dominant component is the helicity $\lambda=s$, for which we find the scaling
\begin{eqnarray}
\Delta_-=1-s-\frac{1}{\a}, ~~~\Delta_+=2+s.
\end{eqnarray}
Repeating the argument for the canonically normalized field 

\be
\bar\sigma_{\mu_1\cdots\mu_s}={\rm exp}(I(\phi_0)/2)\sigma_{\mu_1\cdots\mu_s},
\ee
we find

\be
\bar\Delta_{\pm}-s=\Delta_{\pm}-s+\frac{1}{2\a}.
\ee
Demanding $\bar\Delta_-=0$, we get

\be
\alpha=\frac{1}{2(1-s)}
\ee
and $\bar\Delta_+=3$. Demanding instead $\bar\Delta_+=0$, we get

\be
\alpha=-\frac{1}{2(2+s)}
\ee
which corresponds to  $\bar\Delta_-=3$.

\noindent
\section{Some possible observational consequences}
Long-lived HS fields are active during inflation may give rise to peculiar  signatures on the non-gaussian observed (anisotropic) correlators, which are not necessarily suppressed by the mass of the HS fields or by powers of the long mode in the squeezed limit. Let us discuss in this section some possibilities. 

Consider for instance a massive spin-2 state coupled to the inflation field as described in Section 3 and such that the corresponding scaling at large scales approximately vanishing. The coupling to the comoving curvature perturbation $\zeta$ will be of  the form

\be
S\supset g \int \frac{{\rm d}\tau {\rm d}^3 x}{H^2\tau^2} \,\bar{\sigma}_{ij}\,\partial_i\zeta\,\partial_j\zeta.
\ee
The  exchange of the long-lived  $\pm 2$ polarizations   generates a scalar four-point function in the soft limit $q\equiv | \vec k_1+ \vec k_2| \ll k_i$ ($i=1,\cdots, 4$)

\begin{eqnarray} 
\Big<\zeta_{\vec k_1}\zeta_{\vec k_2}\zeta_{\vec k_3}\zeta_{\vec k_4}\Big>' &\simeq& \Big<\langle \zeta_{\vec k_1}\zeta_{-\vec k_1}\rangle\Big>
 \Big<\langle \zeta_{\vec k_3}\zeta_{-\vec k_3}\rangle\Big>'\nonumber\\
&=&\frac{9}{4} g^2 P_{\bar\sigma}(q) P_\zeta( k_1)P_\zeta(k_3) \sum_{s=\pm 2} \varepsilon^s_{ij}(\vec q) \varepsilon^s_{k\ell}(\vec q) \hat k_{1,i}\hat k_{1,j} \hat k_{3,k} \hat k_{3,\ell},
\end{eqnarray}
where the primes indicate that we have removed the factors $(2\pi)^3$ and the Dirac delta's. This four-point correlator can have a sizable
amplitude and maybe detectable in future CMB and galaxy survey throughout its imprinted anisotropy. In particular, the massless graviton contribution corresponds to $g=1$. Any deviation from it will signal the presence of extra spin-2 states.

Another possible observable where the presence of the spin-2 state might appear is in the power spectrum of the comoving curvature
perturbation. The action might contain terms of the form

\begin{eqnarray}
S&\supset &g_2 H^2 \int \frac{{\rm d}\tau {\rm d}^3 x}{H^4\tau^4} \,{\rm exp}(I)\sigma_{ij}\sigma^{ij}\nonumber\\
&=&g_2 H^2 \int \frac{{\rm d}\tau {\rm d}^3 x}{H^4\tau^4}\,{\rm exp}(I(\phi_0))\left[1+I'(\phi_0)\,\delta\phi+\frac{1}{2}I''(\phi_0)(\delta\phi)^2+\cdots
\right]\sigma_{ij}\sigma^{ij}\nonumber\\
&\simeq&g_2 H^2 \int \frac{{\rm d}\tau {\rm d}^3 x}{H^4\tau^4}\left[1+\frac{1}{\alpha}\,\zeta+\frac{1}{2\alpha^2}\zeta^2+\cdots
\right]\bar{\sigma}_{ij}\bar{\sigma}^{ij}.
\end{eqnarray}
The first term linear in $\zeta$ can be further split if the $\bar{\sigma}^i_{\, j}$ gets an expectation value inside the Hubble radius during inflation. This is expected since, even though such a zero mode is absent at the beginning of inflation, it will be quantum mechanically generated to be of the order of the square root of its variance, $\langle \bar{\sigma}^i_{\, j}\rangle\sim H^2 N$, where $N$ is the total number of $e$-folds. Repeating what done in Ref. \cite{peloso}, one therefore expects
a correction to the power spectrum of the comoving curvature perturbations to be of the order of

\be
\frac{\delta{\cal P}_\zeta(k)}{{\cal P}_\zeta(k)}\sim \frac{g_2^2 H^2 N_k}{\alpha^2\epsilon M_{\rm pl}^2}\sum_{\lambda=\pm 2}\langle \bar{\sigma}^i_{\, j}\rangle\langle \bar{\sigma}^m_{\, \ell}\rangle \varepsilon^j_{\, (\lambda) i}(\hat k)\varepsilon^\ell_{\, (\lambda) m}(\hat k),
\ee
where $N_k$ is the number of $e$-folds to go till the end of inflation from the moment the wavelength $1/k$ leaves the Hubble radius and  $\epsilon=-\dot H/H^2$ is a slow-roll parameter. Parametrizing the anisotropy generated by the helicity-2 background by the 
unit vector $\vec n$

\be
\langle \bar{\sigma}^i_{\, j}\rangle=\langle \bar{\sigma}\rangle(n^in_j-\delta^{i}_{\, j}/3), 
\ee
and exploring the spin sum

\begin{eqnarray}
\sum_{\lambda=\pm 2} \varepsilon^j_{\, (\lambda) i}(\hat k)\varepsilon^\ell_{\, (\lambda) m}(\hat k)&=& 2\left(P^{j\ell}P_{im}+
P_{i}^{\, \ell}P_m^{\, j}\right)-2P_i^{\,j}P_m^{\,\ell},
\nonumber\\
P_{i}^{\, j}&=&\delta_i^{\, j}-\hat{k}_i\hat{k}^j,
\end{eqnarray}
we finally find in terms of the angle $\cos\theta=\vec{n}\cdot\hat k$  

\be
\frac{\delta{\cal P}_\zeta(k)}{{\cal P}_\zeta(k)}\sim  \frac{2g_2^2H^2 N_k \langle \bar{\sigma}\rangle^2}{\alpha^2\epsilon M_{\rm pl}^2}
\sin^4\theta.
\ee
This result can be generalized to a generic spin-$s$
field $\sigma_{\mu_1\cdots \mu_s}$ with an interaction of the form

\begin{eqnarray}
S&\supset &g_s H^2 \int \frac{{\rm d}\tau {\rm d}^3 x}{H^4\tau^4} \,{\rm exp}(I)\sigma_{i_1\cdots i_s}\sigma^{i_1\cdots i_s},
\end{eqnarray}
where $g_s$ a spin dependent coupling. It will lead to a correction to the comoving curvature power spectrum of the form

\be
\frac{\delta{\cal P}_\zeta(k)}{{\cal P}_\zeta(k)}\sim\frac{2g_s^2 H^2 N_k \langle \bar{\sigma}\rangle^2}{\alpha^2\epsilon M_{\rm pl}^2}
\sin^{2s}\theta,
\ee
with a distinctive angle dependence signature.  We see that the presence of HS backgrounds leads to well-defined angular anisotropic structure in the late-time universe observables. In particular, it might interesting to understand if the couplings $g_s$ are related to each other in such a way that the various contributions
can be resummed.  We leave this and other investigations  for the  future.

\noindent
\section{Conclusions}
Inflation offers a unique possibility to  probe high energy states. In this paper we have investigated whether
massive HS fields, if present during inflation, may be quantum mechanically excited and possess fluctuations which are not damped
on super-Hubble scales. While this is not possible in the standard case where HS fields are coupled to the spacetime background minimally, due to the Higuchi bound, we have shown that suitable couplings to functions of the inflaton field may deliver long-lived HS fluctuations on large scales.

Our findings can be generalized in several ways. First, we have restricted ourselves to equations of motions with a maximum of two-derivatives. One could extend the study to higher-derivatives, maybe using the ambient space methods. It might be also worth exploring deformations of the divergence condition we have imposed to reduce the degrees of freedom. Finally, since a consistent theory of HS in de Sitter calls for an infinite tower of fields and 
  although the effect of a single short lived HS field could be observationally negligible, the  the effect of an entire trajectory (non-linearly interacting) could still produce some enhanced effect. 
All these issues clearly stress  the need to construct a consistent theory of HS fields during inflation.

\section*{Acknowledgments}
 We thank C. Sleight and M. Taronna for very useful comments about the draft.
 A.R. is supported by the Swiss National Science Foundation (SNSF), project {\sl Investigating the
Nature of Dark Matter}, project number: 200020-159223.

\appendix
\numberwithin{equation}{section}
\section{Useful relations}
In the text we have made repeatedly use  of the  relation
\begin{eqnarray}
[\nabla_\rho,\nabla_\sigma]\sigma^{\mu_1\cdots\mu_s}={R^{\mu_1}}_{\kappa\rho\sigma}
\sigma^{\kappa\cdots\mu_s}+\cdots +{R^{\mu_s}}_{\kappa\rho\sigma}
\sigma^{\mu_1\cdots\kappa}, \label{com}
\end{eqnarray} 
where $R_{\mu\kappa\rho\sigma}$ is the Riemann tensor.
We also recall here  that a spin-$s$ field has components $\sigma_{\alpha_1\ldots \alpha_s}$ in an  orthonormal local Lorentz frame, where the indices 
$\alpha_1,\ldots, \alpha_s$ are flat. This field transforms in the 
$2s+1$-dimensional representation of the SO(4,1) group of rotations of the orthonormal Lorentz frame. It can written in terms of the totally symmetric
tensor $\sigma^{\mu_1\ldots \mu_s}$ as 
\begin{eqnarray}
\sigma_{\mu_1\ldots \mu_s} &=&e^{\alpha_1}_{\mu_1}\cdots 
e^{\alpha_s}_{\mu_s} \, \sigma^{\mu_1\ldots \mu_s}\sigma_{\alpha_1\ldots \alpha_s}\nonumber\\
&=&\tau^{-s} \, 
\delta^{\alpha_1}_{\mu_1}\cdots \delta^{\alpha_n}_{\mu_n}\, \sigma_{\alpha_1\ldots \alpha_s} , \label{ud}
\end{eqnarray}
where $e^\alpha_\mu=\tau^{-1}\delta^\alpha_\mu$ is the veilbein for the de Sitter metric (\ref{metric})  and $(\mu_1,\ldots, \mu_s)$ are curved space indices.  If $\sigma_{\alpha_1\ldots \alpha_s}$ scales near $\tau\to 0$ as 
\begin{eqnarray}
\sigma_{\alpha_1\ldots \alpha_s}(\t,\vx)\sim \tau^\Delta \hat\sigma_{\alpha_1\ldots \alpha_s}(\vx), ~~~\t\to 0,
\end{eqnarray}
we find that the scaling of $\sigma_{\mu_1\ldots \mu_s} $ is accordingly 
\begin{eqnarray}
\sigma_{\mu_1\ldots \mu_s} (\t,\vx)\sim \tau^{\Delta-s}\hat\sigma_{\mu_1\ldots \mu_s}(\vx),  ~~~\t\to 0.
\end{eqnarray}
Finally, for the  four-dimensional  de Sitter  metric (\ref{metric}), 
the non-vanishing components of the connection are 
\begin{eqnarray}
{\Gamma^k}_{0m}=-\frac{1}{\tau}\delta^k_m, ~~~{\Gamma^0}_{ms}=-\frac{1}{\tau}
\delta_{ms}.
\end{eqnarray}
Correspondingly,   the components of the covariant derivative 
$\nabla_\mu\sigma_{\rho_1\cdots\rho_s}$ are given by
\begin{eqnarray}
&&\nabla_0\sigma_{0\cdots 0 r_{n\!+\!1}\cdots r_s}=\sigma_{0\cdots 0r_{n\!+\!1}\cdots r_s}'+\frac{s-n}{\tau}\ \sigma_{0\cdots 0r_{n\!+\!1}\cdots r_s},\\
&& \nabla_m\sigma_{0\cdots 0 r_{n\!+\!1}\cdots r_s}=\partial_m \sigma_{0\cdots 0 r_{n\!+\!1}\cdots r_s}+\frac{n}{\tau}\,  \sigma_{0\cdots 0 r_nr_{n\!+\!1}\cdots r_s}+\delta_{rr_{n\!+\!1}}\frac{s-n}{\tau}\,
\sigma_{0\cdots 0 r_{n\!+\!2}\cdots r_s}.
\end{eqnarray}


\end{document}